\documentclass[twocolumn]{revtex4}
\usepackage{graphicx}
\usepackage{dcolumn}
\usepackage{longtable}

\begin{document}

\title{Unified description of $0^+$ states in a large class of nuclear collective models}

\author{Dennis Bonatsos$^1$,  E.A. McCutchan$^2$, and R.F. Casten$^3$,}

\affiliation{$^1$Institute of Nuclear Physics, N.C.S.R.
``Demokritos'', GR-15310 Aghia Paraskevi, Attiki, Greece}

\affiliation{$^2$ Physics Division, Argonne National Laboratory,
Argonne, Illinois 60439, USA}

\affiliation{$^3$ Wright Nuclear Structure Laboratory, Yale
University, New Haven, CT 06520, USA}

\begin{abstract}

A remarkably simple regularity in the energies of $0^+$ states in
a broad class of collective models is discussed.  A single formula
for all $0^+$ states in flat-bottomed infinite potentials that
depends only on the number of dimensions and a simpler expression
applicable to all three IBA symmetries in the large $N_B$ limit
are presented. Finally, a connection between the energy expression
for $0^+$ states given by the X(5) model and the predictions of
the IBA near the critical point is explored.

\end{abstract}

\maketitle

The evolution of structure in many-body quantum systems and the
emergence of collective phenomena is a subject that pervades many
areas of modern physics. Recently, significant strides have
been taken in the study of structural evolution in atomic nuclei,
particularly with the discovery of nuclei that undergo
quantum phase transitions~\cite{phase1,phase2} in their
equilibrium shapes, and the development of descriptions of nuclei
at the phase transitional point by simple,
parameter-free critical point symmetries, E(5)~\cite{e5} and
X(5)~\cite{x5}, that invoke flat-bottomed, infinite potentials.
These descriptions have been well supported by experimental studies~\cite{134ba,152sm,150nd,154gd} and also have
application~\cite{molecules} in other systems such as molecules.
Thus their study, and that of related models, potentially
offers insight into a variety of phase transitional behavior.

E(5) and X(5) are analytic solutions of the Bohr
Hamiltonian~\cite{bohr} that describe collective properties in
terms of two shape variables -- the ellipsoidal deformation
$\beta$ and a measure of axial asymmetry, $\gamma$.  Both use
an infinite square well in $\beta$, differing in their $\gamma$
dependence.  Their success has given rise to numerous
other geometrical models, many of which can also be solved
analytically. Examples are X(3)~\cite{x3} in which the $\gamma$ potential
is frozen at $0^{\circ}$, Z(4)~\cite{z4}, in which it is fixed at $\gamma$ = $30^{\circ}$, and Z(5)~\cite{z5} which has a minimum at $\gamma$ = $30^{\circ}$.
In all these models, the number in parentheses is the effective
dimensionality, $D$. For example, the 5-dimensional
models are couched in terms of $\beta$, $\gamma$ and the three
Euler angles.

\begin{figure}
\center{{\includegraphics[height=50mm]{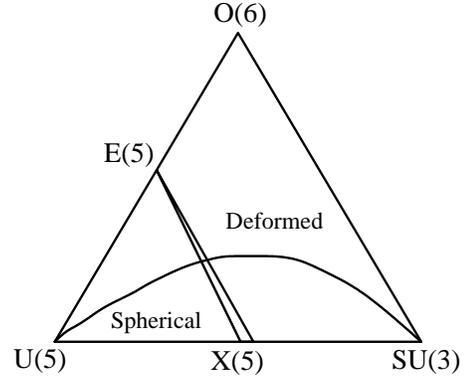}}}
\caption{IBA symmetry triangle with the three dynamical symmetries. The critical point models E(5) and X(5) are placed
close to the phase transition region (slanted
lines). The solid curve indicates the Alhassid-Whelan arc of regularity.}
\end{figure}

The nature of low lying $0^+$ states is critical to understanding
the structure of nuclei. Their identification and interpretation is a subject of recent experimental~\cite{des,dorel} and theoretical
work, both in the microscopic quasi-particle phonon model~\cite{qpm} and in the relativistic mean field framework~\cite{rmf}.

It is the purpose of this Letter to show that a large class of
seemingly diverse models in fact share some remarkable
similarities and to exemplify this by pointing out certain
heretofore unrecognized but simple and general regularities in the
energies of $0^+$ states which characterize these models.  We will
obtain a single, simple formula for {\it all} $0^+$ states in any
flat-bottomed infinite potential that depends {\it solely} on the
number of dimensions and another even simpler expression
applicable to the dynamical symmetries of the Interacting Boson Approximation (IBA) model~\cite{iba} (in the limit of
large valence nucleon number, $N_B$) and compare these results to available data.  We will also use solutions for a series of
potentials with intermediate shapes to study the evolution from
one description (formula) to the other. Finally, we will show that
IBA predictions near the critical point, for large $N_B$, approach
the same energy expression for $0^+$ states as given by the X(5)
critical point symmetry.

In Fig. 1, the IBA triangle is shown with its dynamical symmetries and division into spherical and deformed regimes, separated by a phase transitional region~\cite{phase1}. We include the Alhassid-Whelan(AW) arc of regularity~\cite{arc} and the geometric critical point models E(5) and X(5), although the latter do not belong to the IBA space. 

\begin{figure}
\center{{\includegraphics[height=51mm]{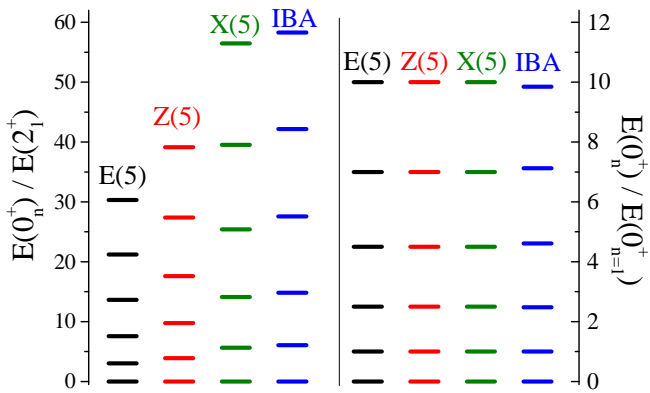}}}
\caption{(Color online) (Left) Energies of excited $0^+$ states in
the E(5), Z(5), and X(5) models as well as an IBA calculation near
the critical point (see text). (Right) Same as left with the
energies normalized to the first excited $0^+$ state energy.}
\end{figure}

For any infinite flat-bottomed (i.e., square well) potential, the
energy eigenvalues are proportional to the squares of roots of the
Bessel functions $J_{\nu}$($z$) where the order $\nu$ is different
for each case.  In E(5)~\cite{e5}, one has
\begin{equation}\label{nE5}
\nu = \tau+{3/2},
\end{equation}
\noindent with $\tau=L/2$. In X(5) one has~\cite{x5}
\begin{equation}\label{nX5}
\nu = \sqrt{ {L(L+1)\over 3} + {9\over 4}}
\end{equation}
\noindent for all $K=0$ bands. In Z(5) one has \cite{z5}
\begin{equation}\label{nZ5}
\nu = {\sqrt{L(L+4)+3 n_w(2L-n_w)+9} \over 2} ,
\end{equation}
\noindent with $n_w=0$ for all $0^+$ bandheads. As a result, in
all three cases, the $0^+$ bandhead energies are given by the
squares of subsequent roots of $J_{3/2}(z)$.

These energies are given in Table I and are plotted on the left in
Fig. 2.  While these seem quite different, if we normalize each
energy to that of the first excited $0^+$ state - that is,
consider relative $0^+$ eigenvalues - we see that these different
models produce exactly identical results as seen on the right in
Fig. 2 and given by (see column Norm in Table I): 0, 1, 2.5, 4.5,
7, 10. These energies are approximately described by the simple formula
\begin{equation}\label{nn3}
E = An(n+3)
\end{equation}
\noindent where $n$ is the ordinal number of the $0^+$ state and
where $A$ depends on the model.

\begin{table}
\caption{Energies of $0^+$ states in the E(5), Z(5), and X(5)
models. Energies on the left are in units of $E$($2_1^+$) = 1.0,
while in the column Norm, in units $E$($0_2^+$) = 1.0. The
normalized results are identical for each of the models. The
column IBA-Norm gives the normalized $0^+$ energies for a large
$N_B$ IBA calculation near the critical point (see text).}
\begin{tabular}{c|c|c|c||c||c}
\hline

$0_i^+$ & E(5) & Z(5) & X(5) & Norm & IBA-Norm\\
\hline
$0_1^+$ & 0 & 0 & 0 & 0  & 0\\
$0_2^+$ & 3.03 & 3.91 & 5.65 & 1.0 & 1.0 \\
$0_3^+$ & 7.58 & 9.78 & 14.12 & 2.50  & 2.48\\
$0_4^+$ & 13.64 & 17.61 & 25.41 & 4.50 & 4.62\\
$0_5^+$ & 21.22 & 27.39 & 39.53 & 7.00 & 7.13\\
$0_6^+$ & 30.31 & 39.12 & 56.47 & 10.00 & 9.85\\
\hline
\end{tabular}
\end{table}

This remarkable result systematizes the $0^+$ states of this wide
array of seemingly diverse models. However, its significance runs
much deeper.  Consider now the models Z(4) in four dimensions and
X(3) in three.

In Z(4) the order of the Bessel functions is given by
\begin{equation}\label{nZ4}
\nu= {\sqrt{L(L+4)+3 n_w (2L-n_w)+4} \over 2},
\end{equation}
\noindent with $n_w=0$ for all $0^+$ states, giving $J_1$($z$) for
$L$=0.

In X(3) one has \cite{x3}
\begin{equation}\label{nX3}
\nu=\sqrt{{L(L+1)\over 3}+{1\over 4}},
\end{equation}
\noindent leading to $J_{1/2}(z)$ for $L=0$.

The $0^+$ energies given by Z(4) and X(3) can again be described
by simple formulas, satisfying
\begin{eqnarray}\label{nn2}
E &=& An(n+2.5)  \;\;\;\;\;\;\;\;\;\;\;\;\;\; \textrm{Z(4)} \nonumber\\
E &=& An(n+2).   \;\;\;\;\;\;\;\;\;\;\;\;\;\;\;\; \textrm{X(3)}
\end{eqnarray}

The results of Eqs. (\ref{nn3}) and (\ref{nn2}) exemplify a
universal formula that depends solely on the number of dimensions.
{\it The energies of all 0+ states in any flat-bottomed infinite
square well potential in D dimensions} are given by:
\begin{equation}\label{map}
E = An\left(n+\frac{D+1}{2}\right)
\end{equation}
\noindent where, again, $A$ depends on the model. Figure 3(a)
shows the results of Eq. (\ref{map}) for $D$ = 3,4, and
5.  These results reflect the deep relation between the order of
the Bessel function solutions and the dimensionality of the
potential, given by $\nu$ = ($D$-2)/2. These results are exact for $D$=3 and excellent approximations for low $\nu$ otherwise. (Compare Eq. (4) with Table II). These findings have applicability well beyond the models
discussed above.  For example, Eq. (\ref{map}) gives the energies
of all $0^+$ states in a recent model~\cite{clark} of the critical
point of a pairing vibration to pairing rotation phase
transition in which the $0^+$ states span two degrees of
freedom -- excitation energies within a given nucleus and the
sequences of masses along a series of even-even nuclei.
Hadronic spectra have also been described~\cite{brodsky} in terms of roots of Bessel functions.

\begin{figure}
\center{{\includegraphics[height=85mm]{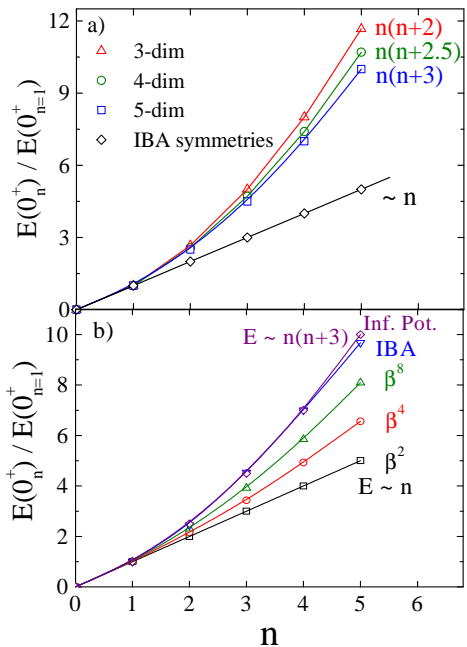}}}
\caption{(Color online) Energies of excited $0^+$ states
(normalized to the first excited $0^+$ state) as a function of
their ordinal number. (a) For infinite square well potentials in 3,4,
and 5 dimensions and for the three IBA dynamical symmetries in the large $N_B$ limit. (b) For potentials of the form
$\beta^{2m}$ as well as for the 5-D square well potential and
an IBA calculation near the critical point (see text). }
\end{figure}

The regularities found for $0^+$ states in solutions of the
Bohr Hamiltonian with an infinite well potential in $\beta$ can
be related to the second order Casimir operator of E($D$), the
Euclidean algebra in $D$ dimensions, which is the semidirect
sum~\cite{Wyb} of the algebra T$_D$ of translations in $D$
dimensions (generated by the momenta),
and the SO($D$) algebra of rotations in $D$ dimensions
(generated by the angular momenta)
 symbolically written as E($D$)=T$_{\rm D}$ $\oplus_s$SO(D)~\cite{Barut}.
The square of the total momentum, $P^2$, is a second order Casimir
operator of the algebra, and eigenfunctions of this operator
satisfy~\cite{z4}
\begin{equation}\label{eq:e56}
 \left( -{1\over r^{D-1}} {\partial \over \partial r} r^{D-1} {\partial \over
\partial r} + { \omega(\omega+D-2) \over r^2} \right) F(r) = k^2 F(r),
\end{equation}
\noindent in the left hand side of which the eigenvalues of the
Casimir operator of SO($D$), $\omega(\omega+D-2)$
appear~\cite{Mosh1555}. Performing the transformation $ F(r) =
r^{(2-D)/2} f(r)$, and using $\nu$ = $\omega$ + $\frac{D-2}{2}$,
Eq. (\ref{eq:e56}) is brought into the form
\begin{equation}\label{eq:e59}
 \left( {\partial^2 \over \partial r^2} + {1\over r} {\partial \over \partial
r} + k^2 - { \nu^2 \over r^2}\right) f(r)  =0,
\end{equation}
\noindent whose solutions are Bessel functions
$f(r)=J_\nu(kr)$.

In the original developments of the infinite square well models,
the ``radial'' equations are obtained, after the above
transformation has been performed, in the form of Eq.
(\ref{eq:e59}) with the corresponding order $\nu$. In E(5) all 
states obey Eq. (\ref{eq:e59}) (with
$\omega=\tau$), in Z(4) agreement occurs for all with $n_w=0$ and
$L=2\omega$~\cite{z4}, while in X(5), X(3) and Z(5) (with $n_w=0$)
agreement is limited to states with $L=0=\omega$, i.e. to the
$0^+$ bandheads.  This situation resembles a partial dynamical
symmetry \cite{AlhLev} of Type I \cite{Lev98}, in which some of
the states (here, the $0^+$ states) preserve all the relevant
symmetry.

\begin{table}
\caption{Experimental $0^+$ levels (in keV) of several nuclei compared to the $0^+_3$ predictions (normalized to the experimental $0_2^+$ energy) of X(5) and SU(3). For SU(3), states belonging to the (2$N$-8,4) and (2$N$-6,0) irreps are given.}
\begin{tabular}{ccc|ccc}
\hline
&   Exp. $0^+$  &  $\; $  X(5) $0_3^+$ & & Exp. $0^+$ &  $\;$  SU(3) $0_3^+$ \\
$^{150}$Nd & 1738 & 1688 & $^{156}$Gd & 1851  & 1916 \\
$^{152}$Sm & 1659 & 1712 &            & 1989  & 2054 \\
$^{154}$Gd & 1650 & 1702 & $^{158}$Gd & 2276  & 2201 \\
           &      &      &            & 2338  & 2344  \\

\hline

\end{tabular}
\end{table}

Do we find these patterns in real nuclei? As examples, we use the well-studied X(5) candidates, $^{150}$Nd, $^{152}$Sm and $^{154}$Gd. Normalizing to the experimental $0_2^+$ energy in each nucleus, Table II gives the $0^+_3$ energies predicted by Eqs. (4) and (8).  In each nucleus, there is indeed a $0^+$ state within $<$ 100 keV of the predicted energy.  In $^{150}$Nd this is in fact the $0_3^+$ state, while in the rest, it corresponds to a higher lying $0^+$ state, for which the determination of the degree of collectivity through improved spectroscopic information poses an experimental challenge.

Eqs. (\ref{nn3}) and (\ref{nn2}) are peculiar to
infinite square wells and thus limited to a select number of nuclei. 
What behavior then characterizes other potentials? Of course, this cannot 
be solved in general but there is at least
one other class of models where an easy solution can be derived,
namely, in the dynamical symmetries of the IBA.

In U(5), the $0^+$ energies are simply proportional 
to the number of $d$ bosons defining their respective phonon number, thus
$E$($0^+_n$) = $An$. In SU(3), the eigenvalue expression 
for $L$ = $0^+$ states, in
terms of the usual representation labels ( $\lambda$, $\mu$), is
$E$ = $a$[$\lambda^2$ + $\mu^2$ + $\lambda \mu$ + 3($\lambda$ +
$\mu)$].  In O(6), the corresponding equation, in terms of the
major family quantum number $\sigma$, is (for $\tau$ =0,$L=0$
states) $E$ = $a$ $\sigma$($\sigma$+4). 
The irreducible representations for SU(3) and O(6) and the
corresponding $0^+$ energies are given in Table III. Taking
successive ($\lambda$, $\mu$) and $\sigma$ quantum numbers and the
limit $N_B$$\rightarrow$$\infty$ we obtain:
\begin{equation}\label{n}
E = An .
\end{equation}
Thus, perhaps surprisingly, considering how different their
structures are, and analogous to the infinite flat potentials, a
single, simple formula applies to {\it all three dynamical
symmetries} of the IBA which exhibit identical relative energy
spectra of $0^+$ states in the large $N_B$ limit. Equations
(\ref{map}) and (\ref{n}) are compared in Fig. 3(a).

\begin{table}
\caption{Irreducible representations (irreps) of SU(3) and O(6) and
the corresponding energy of the excited $0^+$ states.}
\begin{tabular}{cc|cc}
\hline
\multicolumn{2}{c}{SU(3)} & \multicolumn{2}{c}{O(6)}\\
\hline

Irrep ($\lambda$,$\mu$) & $E$($0^+$) & Irrep ($\sigma$) & $E$($0^+$) \\

\hline

(2$N$,0) & 0 & ($N$) & 0 \\
(2$N$-4,2) & 1 & ($N$-2) & 1 \\
(2$N$-8,4) & (4$N$-6)/(2$N$-1) & ($N$-4) & 2\\
(2$N$-6,0) & (4$N$-3)/(2$N$-1) & ($N$-6) & 3-(3/$N$)\\
(2$N$-12,6) & (6$N$-15)/(2$N$-1) & ($N$-8) & 4-(8/$N$)\\

\hline
\end{tabular}
\end{table}

The description of $0^+$ states with a simple analytic 
formula extends beyond just the vertices discussed above. 
The behavior of $0^+$ states in the IBA symmetries can be associated with the chaotic properties of the IBA~\cite{arc}.  A regular region connects U(5) and O(6), resulting from the underlying O(5) symmetry. The regular behavior associated with U(5) and O(6) is preserved along the U(5)-O(6) leg, manifesting the relevant quasidynamical symmetries~\cite{rowea}, until close to the point of the second order transition.  An additional regular region, the AW arc of regularity, connects U(5) and SU(3) through the interior of the triangle (see Fig. 1).  It has been conjectured~\cite{chaos} that the regular region between SU(3) and the critical line is related to an underlying partial SU(3) symmetry.  Included in Table II is a comparison between two well-deformed nuclei proposed~\cite{jolie} to lie along the arc of regularity with the SU(3) predictions of Table III for the $0_3^+$ state and its nearly degenerate companion.  Good agreement is observed, although the degree of collectivity of these $0^+$ states needs further experimental examination.

Most nuclei, however, are not described by dynamical
symmetries or critical point models but lie somewhere in
between~\cite{mapping}.  To study this and to see how Eq.
(\ref{nn3}) and Eq. (\ref{n}) are
related, or, better, how they evolve into one another moving
across the symmetry triangle, we now consider a sequence of
potentials of the form, V $\sim$ $\beta^{2m}$ starting from
$\beta^2$, corresponding to U(5), and ending at either X(5) or
E(5), which is successively approached by increasing powers
of $\beta$. The results are shown in Fig 3(b).  As the potential
flattens with increasing powers of $\beta$, the results go from
those for the U(5) limit to those for the infinite
square well potentials.  In each case, the normalized $0^+$
energies are well reproduced by a formula analogous to Eq.
(\ref{nn3}), namely $E$ $\sim$ $n$($n$+$x$) where
$x$$\rightarrow$$\infty$ for the IBA symmetries and drops to 3 for
E(5)/X(5).  

Finally, it has recently been discussed how the IBA with
appropriate parameters approaches the predictions of the critical
point symmetries as $N_B$$\rightarrow$$\infty$~\cite{newPRL}. We use
an IBA Hamiltonian in the form~\cite{Werner}
\begin{equation}\label{ham}
H(\zeta,\chi) = c \left[ (1-\zeta) \hat n_d -{\zeta\over 4 N_B}
\hat Q^\chi \cdot \hat Q^\chi\right],
\end{equation}
\noindent where $\hat n_d = d^\dagger \cdot \tilde d$, $\hat
Q^\chi = (s^\dagger \tilde d + d^\dagger s) +\chi (d^\dagger
\tilde d)^{(2)},$ $N_B$ is the number of valence bosons, and $c$
is a scaling factor. Calculations were performed with the IBAR
code~\cite{ibar}.

Included in Fig. 3(b) is an IBA calculation with  $\chi$ = -$\sqrt{7}$/2
(bottom leg of the triangle), $N_B$ = 250 and
$\zeta$ = 0.473~\cite{newPRL}, which is very close to the critical
point ($\zeta_{crit}$ = 0.472) of the phase transition
region in the IBA. One sees that the IBA results are
very close to those of Eq. (\ref{nn3}) consistent with the flat
nature of the IBA energy surface near the critical point. This
same result is also illustrated in Fig. 2. The normalized IBA energies
are included in Table I and show a strong
similarity with the results of the infinite square well
potentials. Thus, it appears that the regularities in $0^+$
energies obtained in the infinite square well potentials are not
restricted to geometrical models and also occur near the
critical point of the IBA.

In summary, we have discussed very simple regularities in excited
$0^+$ energies which pervade a number of different models. The energies of excited $0^+$ states in flat-bottomed
infinite potentials can be described by a single expression
dependent on only the number of dimensions. These observed
regularities in $0^+$ energies are linked to the second order
Casimir operator of E($n$).  Further, the energies of
$0^+$ states in all three dynamical symmetries of the IBA are
governed, in the large $N_B$ limit, by a single expression.
These results were compared to experimental data.  
Potentials with shapes intermediate between
the U(5) symmetry and the infinite square well give a
smooth evolution in the $0^+$ energies.  Finally, IBA calculations with large $N_B$, near
the critical point of the phase transition, exhibit nearly the
same energy dependence for $0^+$ states as given by the infinite
square well potentials.

The authors thank R.J. Casperson, E. Williams, and V. Werner for
their expertise with the IBAR code and R.M. Clark and A.O.
Macchiavelli for useful discussions. This work was supported by
U.S. DOE Grant No. DE-FG02-91ER-40609 and by the DOE Office of
Nuclear Physics under contract DE-AC02-06CH11357.

\end{document}